\documentclass[a4paper,11pt]{article}
\usepackage{pos}
\usepackage{tikz}
\usepackage{float}
\usepackage{subfigure} 

\title{Topology in 2D non-Abelian Lattice Gauge Theories}

\author*[a]{P. Rouenhoff}
\author[a,b]{S. Dürr}

\affiliation[a]{Physics Department, University of Wuppertal, 42119 Wuppertal, Germany}
\affiliation[b]{IAS/JSC, Forschungszentrum Jülich, 52425 Jülich, Germany}

\emailAdd{p.rouenhoff$\,(\text{AT})\,$uni-wuppertal$\mbox{.}$de}
\emailAdd{durr$\,(\text{AT})\,$itp$\mbox{.}$unibe$\mbox{.}$ch}

\abstract{
    In two dimensions, $U(N_c)$ gauge theories exhibit a non-trivial topological structure, while $SU(N_c)$ theories are topologically trivial. Hence, for $G = U(N_c)$ the phase space is divided into topological sectors, characterized by a topological index (a.k.a. ``topological charge''). These sectors are separated by action barriers, which diverge if the lattice spacing is taken small, resulting in an algorithmic problem known as ``topological freezing''. We study these theories in various box sizes and at various couplings. With the help of gradient flow we derive instanton-like solutions for 2D $U(N_c)$ theory with a specific focus on the case of $N_c = 2$.
}

\FullConference{The 41st International Symposium on Lattice Field Theory (LATTICE2024)\\
 28 July - 3 August 2024\\
Liverpool, UK\\}

\begin{document}
\maketitle

\section{Introduction} 

The phase spaces of two-dimensional $SU(N_c)$ lattice gauge theories on the torus exhibit a trivial topological structure. 
This changes when the gauge group is replaced by $U(N_c)$, which is due to the $U(1)$-factor in $U(N_c) \simeq \frac{U(1) \times SU(N_c)}{\mathbb{Z}_2}$ with the cyclic group $\mathbb{Z}_2$ \cite{Baig_1986, Roiesnel_1995, Hall}.
In that case the phase space is divided into topological sectors, regions in which the topological charge $q$, a topological index that can be assigned to a gauge configuration, is constant. 
For each $q$ one can find a field configuration that minimizes the gauge action, which is called an instanton.
Hence, between these sectors, there are action barriers. 
As they diverge with vanishing lattice spacing $a$, an algorithmic problem known as ``topological freezing'' arises, which affects both 2D $U(N_c)$ as well as 4D $SU(N_c)$ gauge theory.

\begin{figure}[H]
    \centering
    \subfigure{
        \includegraphics[width=0.5\linewidth]{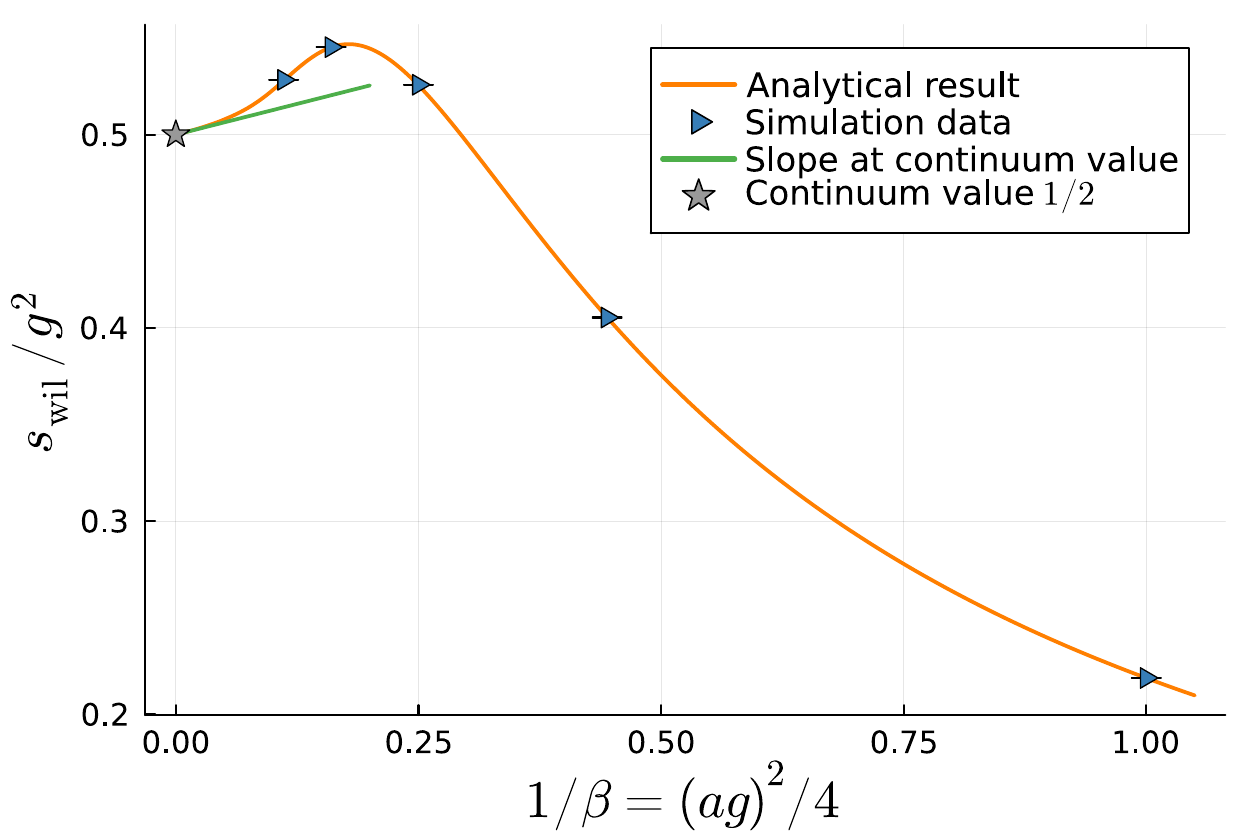}
        \label{fig:s_wil}
    }%
    \subfigure{
        \includegraphics[width=0.5\linewidth]{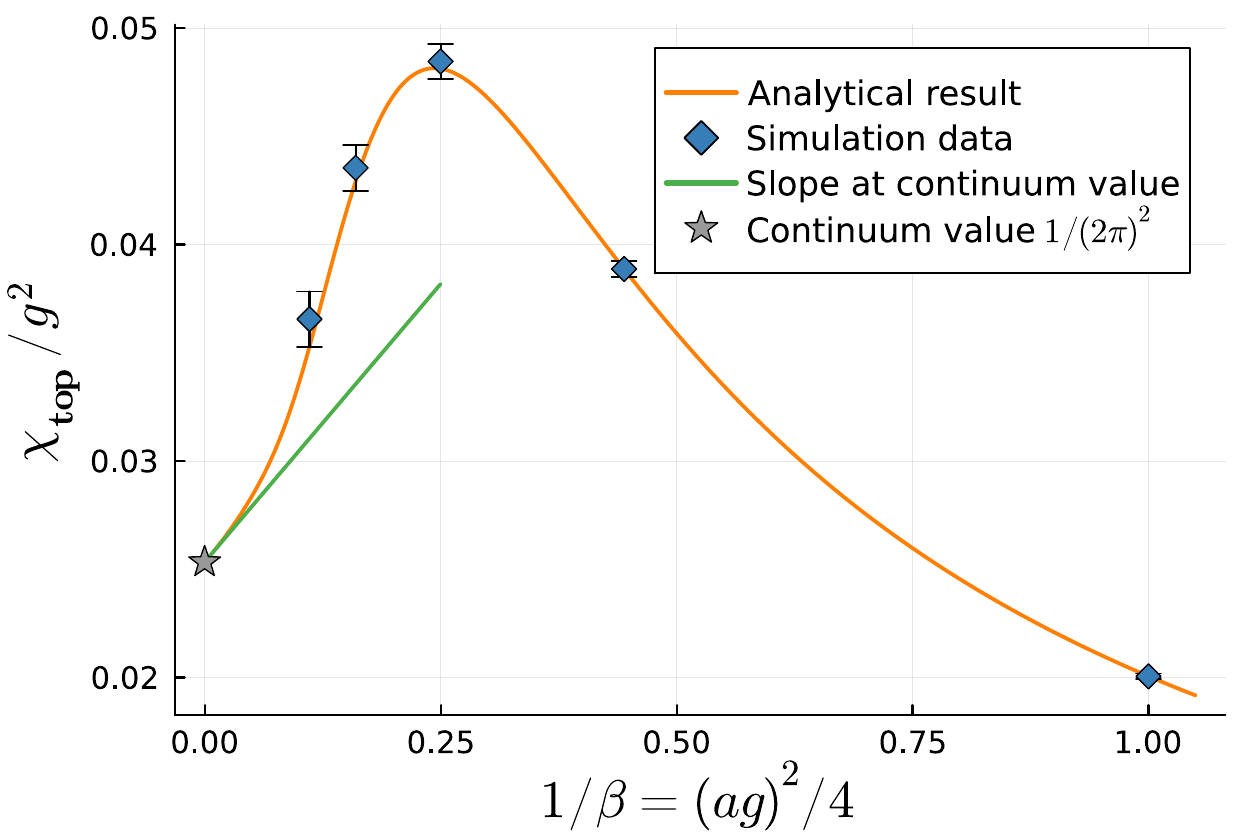}
        \label{fig:susc}
    }%
    \caption{The action density and topological susceptibility of 2D $U(2)$ theory as given in Tab.\,\ref{tab:susc}. In the limit $\beta\rightarrow\infty$ we find that $s_\mathrm{wil}/g^2 \rightarrow 1/2$\ and $\chi_\mathrm{top}/g^2 \rightarrow 0.02533 = 1/(2\pi)^2$. }
    \label{fig:s_wil_susc}
\end{figure}

In order to gain insight into this problem, let us examine the topological structure of the phase space more closely. 
An integer-valued definition for $q$ in 2D $U(N_c)$ theory is given as \cite{Bonati_2019, Bonati_2019_2, Hirasawa_2021}
\begin{equation}\label{eq:q}
     q \equiv \frac{1}{2\pi} \sum_{n\in\Lambda} \mathrm{Im}\left[ \log \det U_\Box(n) \right]
\end{equation}
with the $U(N_c)$-valued (untraced) plaquette
\begin{equation}
    U_\Box(n) = U_x(n) U_t(n+\hat{x}) U_x^\dagger(n+\hat{t}\, ) U_t^\dagger(n),
\end{equation}
which is also needed for the Wilson gauge action which we use throughout this work:
\begin{equation}
    S[U] = \frac{\beta}{N_c} \sum_{n\in\Lambda} \mathrm{Re\, Tr} \big(\mathbf{1}-U_\Box(n)\big).
\end{equation}
A handy parametrization of $U(2)$ via $SU(2) \simeq S^3$ allows to calculate the gauge action as
\begin{equation}\label{eq:s_wil}
    \frac{1}{2}\,\langle \mathrm{Re\, Tr}\, U_\Box \rangle^{U(2)} = \frac{\int_0^{\frac{\pi}{2}} \mathrm{d}\alpha \ \Big[I_0(\beta\cos(\alpha)) + I_2(\beta\cos(\alpha)) \Big] }{2 \, \int_0^{\frac{\pi}{2}} \mathrm{d}\alpha \  \frac{I_1(\beta\cos(\alpha))}{\cos(\alpha)}} -\frac{1}{\beta}\ ,
\end{equation}
with the 2D convention $\beta \equiv \frac{2N_c}{a^2g^2}$. 
Similarly the topological susceptibility, defined as $\chi_\text{top} \equiv \frac{\langle q^2\rangle}{V}$, is
\begin{equation}\label{eq:susc}
    a^2 \chi_\text{top}^{U(2)} = \frac{\int_0^{\frac{\pi}{2}} \mathrm{d}\alpha \ \alpha^2 \,  \frac{I_1(\beta\cos(\alpha))}{\cos(\alpha)}}{\pi^2 \int_0^{\frac{\pi}{2}} \mathrm{d}\alpha \  \frac{I_1(\beta\cos(\alpha))}{\cos(\alpha)}}\ .
\end{equation}
We compare measurements of $a^2 \chi_\text{top}^{U(2)}$ and $\langle \mathrm{Re\, Tr}\, U_\Box \rangle^{U(2)}$ to the analytic results in Tab.\,\ref{tab:susc} and Fig.\,\ref{fig:s_wil_susc}.
For further analytical results in 2D $U(N_c)$ theory see Refs.\,\cite{Bonati_2019, Bonati_2019_2}.

\begin{table}[H]
    \centering
    \begin{tabular}{c c|c c |c c} 
        $\beta$ &   $L/a$ & $s_\mathrm{wil}/g^2$ &  analyt. value \eqref{eq:s_wil} & $\chi_\text{top}/g^2$ & analyt. value \eqref{eq:susc} \\\hline
        1.0 &   16 &    0.218802(78) & 0.21875985 &     0.02005(12) &  0.02006011   \\
        2.25 &  24 &    0.40535(12) &  0.40542353 &     0.03887(36) &  0.03886658   \\
        4.0 &   32 &    0.52582(21) &  0.52555107 &     0.04846(79) &  0.04812073   \\
        6.25 &  40 &    0.54526(26) &  0.54522256 &     0.0435(10) &   0.04291103   \\
        9.0 &   48 &    0.52835(21) &  0.52794370 &     0.0366(12) &   0.03527802   \\
    \end{tabular}
    \caption{Measurements of the Wilson action density resp. topological susceptibility of 2D $U(2)$ theory for various $(\beta,L/a)$-pairs which implement a fixed physical box volume, compared to the analytical solution.} 
    \label{tab:susc}
\end{table}

\section{Global minima per topological sector}\label{sec:global}

In 2D $U(1)$ theory formulas for instanton configurations, i.e. field configurations of minimal action and fixed topological charge $q\in\mathbb{Z}$, are known as \cite{Smit_1986} 
\begin{align}\label{eq:insta_U1}
    U_x(x,t) = \mathrm{e}^{-\mathrm{i} t \frac{2\pi q}{N_x N_t}}, \quad 
    U_t(x,t) = \mathrm{e}^{\mathrm{i} x \frac{2\pi q}{N_x}\,\delta_{t,N_t}},
\end{align}
with $S^{U(1)}_\mathrm{inst} = \beta N_x N_t \left(\frac{2\pi q}{N_x N_t} \right)$.
In 2D $U(2)$ theory, instanton configurations can be derived as
\begin{align}\label{eq:insta_U2}
    U_x(x,t) = \mathrm{e}^{-\mathrm{i}t \frac{\pi q}{N_x N_t}} \exp(\mathrm{i}\vec{u} \Vec{\sigma}\,\delta_{x,N_x} ), \quad \, 
    U_t(x,t) = \mathrm{e}^{\mathrm{i}x \frac{\pi q}{N_x}\,\delta_{t,N_t}} \exp(\mathrm{i}\vec{v} \Vec{\sigma}\,\delta_{t,N_t} )\ ,
\end{align}
where $\vec{\sigma}$ contains the Pauli matrices and the auxiliary vectors $\vec{u}, \vec{v} \in \mathbb{R}^3$ are subject to the following constraints. 
If $q$ is odd, require that $\Vec{u} \perp \Vec{v}$ and $|\vec{u}| = |\vec{v}| = \frac{\pi}{2}$, and if $q$ is even,  require only $\Vec{u} \parallel \Vec{v}$.

\begin{figure}[!b]
\centering
\resizebox{0.99\textwidth}{!}{%
\begin{tikzpicture}
\tikzstyle{every node}=[font=\huge]
\draw [ color={rgb,255:red,55; green,126; blue,184}, line width=2pt] (3.75,12.75) -- (3.75,15.25); 
\draw [ color={rgb,255:red,55; green,126; blue,184}, line width=2pt] (6.25,12.75) -- (6.25,13.45); 
\draw [ color={rgb,255:red,55; green,126; blue,184}, line width=2pt] (6.25,14.75) -- (6.25,15.25); 
\draw [ color={rgb,255:red,55; green,126; blue,184}, line width=2pt] (8.75,12.75) -- (8.75,13.3); 
\draw [ color={rgb,255:red,55; green,126; blue,184}, line width=2pt] (8.75,14.65) -- (8.75,15.25); 
\draw [ color={rgb,255:red,55; green,126; blue,184}, line width=2pt] (11.25,12.75) -- (11.25,15.25); 
\draw [ color={rgb,255:red,228; green,26; blue,28}, line width=2pt] (11.25,12.75) -- (13.75,12.75); 
\draw [ color={rgb,255:red,228; green,26; blue,28}, line width=2pt] (11.25,10.25) -- (13.75,10.25);
\draw [ color={rgb,255:red,228; green,26; blue,28}, line width=2pt] (11.25,7.75) -- (13.75,7.75);
\draw [ color={rgb,255:red,228; green,26; blue,28}, line width=2pt] (11.25,5.25) -- (13.75,5.25);
\draw [ color={rgb,255:red,153; green,153; blue,153}, line width=2pt] (3.75,5.25) -- (11.25,5.25);
\draw [ color={rgb,255:red,153; green,153; blue,153}, line width=2pt] (3.75,7.75) -- (11.25,7.75);
\draw [ color={rgb,255:red,153; green,153; blue,153}, line width=2pt] (3.75,10.25) -- (11.25,10.25);
\draw [ color={rgb,255:red,153; green,153; blue,153}, line width=2pt] (3.75,12.75) -- (11.25,12.75);
\draw [ color={rgb,255:red,153; green,153; blue,153}, line width=2pt, dashed] (3.75,5.25) -- (3.75,12.75);
\draw [ color={rgb,255:red,153; green,153; blue,153}, line width=2pt, dashed] (6.25,5.25) -- (6.25,12.75);
\draw [ color={rgb,255:red,153; green,153; blue,153}, line width=2pt, dashed] (8.75,5.25) -- (8.75,12.75);
\draw [ color={rgb,255:red,153; green,153; blue,153}, line width=2pt, dashed] (11.25,5.25) -- (11.25,8.25); 
\draw [ color={rgb,255:red,153; green,153; blue,153}, line width=2pt, dashed] (11.25,9.3) -- (11.25,12.75);
\draw [ fill={rgb,255:red,222; green,222; blue,222} , line width=0.7pt ] (3.75,5.25) circle (0.25cm);
\draw [ fill={rgb,255:red,222; green,222; blue,222} , line width=0.7pt ] (6.25,5.25) circle (0.25cm);
\draw [ fill={rgb,255:red,222; green,222; blue,222} , line width=0.7pt ] (8.75,5.25) circle (0.25cm);
\draw [ fill={rgb,255:red,222; green,222; blue,222} , line width=0.7pt ] (11.25,5.25) circle (0.25cm);
\draw [ fill={rgb,255:red,222; green,222; blue,222} , line width=0.7pt ] (3.75,7.75) circle (0.25cm);
\draw [ fill={rgb,255:red,222; green,222; blue,222} , line width=0.7pt ] (6.25,7.75) circle (0.25cm);
\draw [ fill={rgb,255:red,222; green,222; blue,222} , line width=0.7pt ] (8.75,7.75) circle (0.25cm);
\draw [ fill={rgb,255:red,222; green,222; blue,222} , line width=0.7pt ] (11.25,7.75) circle (0.25cm);
\draw [ fill={rgb,255:red,222; green,222; blue,222} , line width=0.7pt ] (3.75,10.25) circle (0.25cm);
\draw [ fill={rgb,255:red,222; green,222; blue,222} , line width=0.7pt ] (6.25,10.25) circle (0.25cm);
\draw [ fill={rgb,255:red,222; green,222; blue,222} , line width=0.7pt ] (8.75,10.25) circle (0.25cm);
\draw [ fill={rgb,255:red,222; green,222; blue,222} , line width=0.7pt ] (11.25,10.25) circle (0.25cm);
\draw [ fill={rgb,255:red,222; green,222; blue,222} , line width=0.7pt ] (3.75,12.75) circle (0.25cm);
\draw [ fill={rgb,255:red,222; green,222; blue,222} , line width=0.7pt ] (6.25,12.75) circle (0.25cm);
\draw [ fill={rgb,255:red,222; green,222; blue,222} , line width=0.7pt ] (8.75,12.75) circle (0.25cm);
\draw [ fill={rgb,255:red,222; green,222; blue,222} , line width=0.7pt ] (11.25,12.75) circle (0.25cm);
\node [font=\huge, color={rgb,255:red,110; green,110; blue,110}] at (5,4.5) {$\mathrm{e}^{-\mathrm{i}t \frac{\pi q}{N_x N_t}}$};
\node [font=\huge, color={rgb,255:red,110; green,110; blue,110}] at (3.25,6.5) {$\mathbf{1}$};
\node [font=\huge, color={rgb,255:red,228; green,26; blue,28}] at (13.0,8.5) {$\mathrm{e}^{-\mathrm{i}t \frac{\pi q}{N_x N_t}} \exp(\mathrm{i}\vec{u} \Vec{\sigma})$};
\node [font=\huge, color={rgb,255:red,55; green,126; blue,184}] at (7.5,14.05) {$\mathrm{e}^{\mathrm{i}x \frac{\pi q}{N_x}} \exp(\mathrm{i}\vec{v} \Vec{\sigma})$};
\draw [ fill={rgb,255:red,255; green,240; blue,240} , line width=1pt , rounded corners = 20.4] (11.5,15) rectangle  node {\huge $\mathrm{e}^{\mathrm{i}\frac{\pi q}{N_x N_t}}$} (13.5,13);
\draw [line width=1pt, ->] (12.25,13) -- (12.5,13);
\draw [line width=1pt, ->] (12.75,15) -- (12.5,15);
\draw [ fill={rgb,255:red,255; green,240; blue,240} , line width=1pt , rounded corners = 20.4] (4,10) rectangle  node {\huge $\mathrm{e}^{\mathrm{i}\frac{\pi q}{N_x N_t}}$} (6,8);
\draw [line width=1pt, ->] (4.75,8) -- (5,8);
\draw [line width=1pt, ->] (5.25,10) -- (5,10);

\draw [ color={rgb,255:red,55; green,126; blue,184}, line width=2pt] (13.50+3.75,12.75) -- (13.50+3.75,15.25); 
\draw [ color={rgb,255:red,55; green,126; blue,184}, line width=2pt] (13.50+6.25,12.75) -- (13.50+6.25,13.45); 
\draw [ color={rgb,255:red,55; green,126; blue,184}, line width=2pt] (13.50+6.25,14.75) -- (13.50+6.25,15.25); 
\draw [ color={rgb,255:red,55; green,126; blue,184}, line width=2pt] (13.50+8.75,12.75) -- (13.50+8.75,13.3); 
\draw [ color={rgb,255:red,55; green,126; blue,184}, line width=2pt] (13.50+8.75,14.65) -- (13.50+8.75,15.25); 
\draw [ color={rgb,255:red,55; green,126; blue,184}, line width=2pt] (13.50+11.25,12.75) -- (13.50+11.25,15.25); 
\draw [ color={rgb,255:red,255; green,127; blue,0}, line width=2pt] (13.50+11.25,12.75) -- (13.50+13.75,12.75); 
\draw [ color={rgb,255:red,255; green,127; blue,0}, line width=2pt] (13.50+11.25,10.25) -- (13.50+13.75,10.25);
\draw [ color={rgb,255:red,255; green,127; blue,0}, line width=2pt] (13.50+11.25,7.75) -- (13.50+13.75,7.75);
\draw [ color={rgb,255:red,255; green,127; blue,0}, line width=2pt] (13.50+11.25,5.25) -- (13.50+13.75,5.25);
\draw [ color={rgb,255:red,153; green,153; blue,153}, line width=2pt, dashed] (13.50+3.75,5.25) -- (13.50+11.25,5.25);
\draw [ color={rgb,255:red,153; green,153; blue,153}, line width=2pt] (13.50+3.75,7.75) -- (13.50+11.25,7.75);
\draw [ color={rgb,255:red,153; green,153; blue,153}, line width=2pt] (13.50+3.75,10.25) -- (13.50+11.25,10.25);
\draw [ color={rgb,255:red,153; green,153; blue,153}, line width=2pt] (13.50+3.75,12.75) -- (13.50+11.25,12.75);
\draw [ color={rgb,255:red,153; green,153; blue,153}, line width=2pt, dashed] (13.50+3.75,5.25) -- (13.50+3.75,12.75);
\draw [ color={rgb,255:red,153; green,153; blue,153}, line width=2pt, dashed] (13.50+6.25,5.25) -- (13.50+6.25,6.35);
\draw [ color={rgb,255:red,153; green,153; blue,153}, line width=2pt, dashed] (13.50+6.25,7.1) -- (13.50+6.25,12.75);
\draw [ color={rgb,255:red,153; green,153; blue,153}, line width=2pt, dashed] (13.50+8.75,5.25) -- (13.50+8.75,6.5);
\draw [ color={rgb,255:red,153; green,153; blue,153}, line width=2pt, dashed] (13.50+8.75,7.4) -- (13.50+8.75,12.75);
\draw [ color={rgb,255:red,153; green,153; blue,153}, line width=2pt, dashed] (13.50+11.25,5.25) -- (13.50+11.25,8.3); 
\draw [ color={rgb,255:red,153; green,153; blue,153}, line width=2pt, dashed] (13.50+11.25,9.3) -- (13.50+11.25,12.75);
\draw [ fill={rgb,255:red,222; green,222; blue,222} , line width=0.7pt ] (13.50+3.75,5.25) circle (0.25cm);
\draw [ fill={rgb,255:red,222; green,222; blue,222} , line width=0.7pt ] (13.50+6.25,5.25) circle (0.25cm);
\draw [ fill={rgb,255:red,222; green,222; blue,222} , line width=0.7pt ] (13.50+8.75,5.25) circle (0.25cm);
\draw [ fill={rgb,255:red,222; green,222; blue,222} , line width=0.7pt ] (13.50+11.25,5.25) circle (0.25cm);
\draw [ fill={rgb,255:red,222; green,222; blue,222} , line width=0.7pt ] (13.50+3.75,7.75) circle (0.25cm);
\draw [ fill={rgb,255:red,222; green,222; blue,222} , line width=0.7pt ] (13.50+6.25,7.75) circle (0.25cm);
\draw [ fill={rgb,255:red,222; green,222; blue,222} , line width=0.7pt ] (13.50+8.75,7.75) circle (0.25cm);
\draw [ fill={rgb,255:red,222; green,222; blue,222} , line width=0.7pt ] (13.50+11.25,7.75) circle (0.25cm);
\draw [ fill={rgb,255:red,222; green,222; blue,222} , line width=0.7pt ] (13.50+3.75,10.25) circle (0.25cm);
\draw [ fill={rgb,255:red,222; green,222; blue,222} , line width=0.7pt ] (13.50+6.25,10.25) circle (0.25cm);
\draw [ fill={rgb,255:red,222; green,222; blue,222} , line width=0.7pt ] (13.50+8.75,10.25) circle (0.25cm);
\draw [ fill={rgb,255:red,222; green,222; blue,222} , line width=0.7pt ] (13.50+11.25,10.25) circle (0.25cm);
\draw [ fill={rgb,255:red,222; green,222; blue,222} , line width=0.7pt ] (13.50+3.75,12.75) circle (0.25cm);
\draw [ fill={rgb,255:red,222; green,222; blue,222} , line width=0.7pt ] (13.50+6.25,12.75) circle (0.25cm);
\draw [ fill={rgb,255:red,222; green,222; blue,222} , line width=0.7pt ] (13.50+8.75,12.75) circle (0.25cm);
\draw [ fill={rgb,255:red,222; green,222; blue,222} , line width=0.7pt ] (13.50+11.25,12.75) circle (0.25cm);
\node [font=\huge, color={rgb,255:red,110; green,110; blue,110}] at (13.50+7.6,6.9) {$\mathrm{e}^{-\mathrm{i}(t+x) \frac{\pi q}{N_x N_t}}$};
\node [font=\huge, color={rgb,255:red,110; green,110; blue,110}] at (13.50+5,4.5) {$\mathbf{1}$};
\node [font=\huge, color={rgb,255:red,110; green,110; blue,110}] at (13.50+3.25,6.5) {$\mathbf{1}$};
\node [font=\huge, color={rgb,255:red,255; green,127; blue,0}] at (13.50+14,8.6) {$\mathrm{e}^{-\mathrm{i}(t+N_x-1) \frac{\pi q}{N_x N_t}} \exp(\mathrm{i}\vec{u} \Vec{\sigma})$};
\node [font=\huge, color={rgb,255:red,55; green,126; blue,184}] at (13.50+7.5,14.05) {$\mathrm{e}^{\mathrm{i}x \frac{\pi q}{N_x}} \exp(\mathrm{i}\vec{v} \Vec{\sigma})$};
\draw [ fill={rgb,255:red,255; green,240; blue,240} , line width=1pt , rounded corners = 20.4] (13.50+11.5,15) rectangle  node {\huge $\mathrm{e}^{\mathrm{i}\frac{\pi q}{N_x N_t}}$} (13.50+13.5,13);
\draw [line width=1pt, ->] (13.50+12.25,13) -- (13.50+12.5,13);
\draw [line width=1pt, ->] (13.50+12.75,15) -- (13.50+12.5,15);
\draw [ fill={rgb,255:red,255; green,240; blue,240} , line width=1pt , rounded corners = 20.4] (13.50+4,10) rectangle  node {\huge $\mathrm{e}^{\mathrm{i}\frac{\pi q}{N_x N_t}}$} (13.50+6,8);
\draw [line width=1pt, ->] (13.50+4.75,8) -- (13.50+5,8);
\draw [line width=1pt, ->] (13.50+5.25,10) -- (13.50+5,10);
\end{tikzpicture}
}
\caption{Left: A visualization of \eqref{eq:insta_U2}, the instanton-like solution for 2D $U(2)$ theory with topological charge $q$. Right: the same configuration in maximal tree gauge. Unity matrices are shown as dashed lines. }
\label{fig:insta_sketch}
\end{figure}
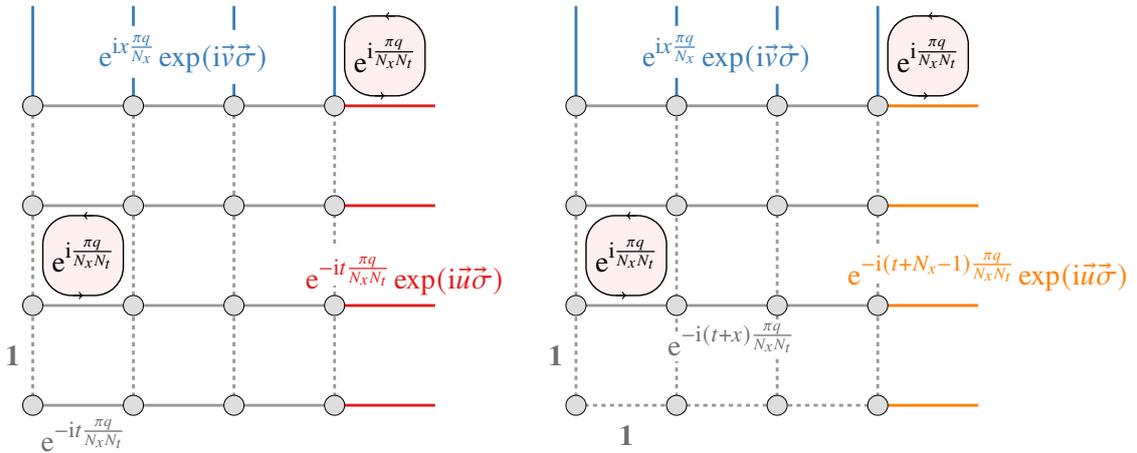

This construction is illustrated in Fig.\,\ref{fig:insta_sketch} for two different gauges.
All horizontal links take the value $\mathrm{e}^{-\mathrm{i}t\frac{\pi q}{N_x N_t}}$, save for those at $x = N_x$. 
This last $x$-slice receives an additional factor of $\exp(\mathrm{i}\vec{u}\vec{\sigma})$.
All vertical links are unity $\mathbf{1}$, except for those at $t = N_t$, which are $\mathrm{e}^{\mathrm{i}x \frac{\pi q}{N_x}} \exp(\mathrm{i}\vec{v} \Vec{\sigma})$.
Despite the four cases implied in this construction, each untraced plaquette takes the same value $\mathrm{e}^{\mathrm{i}\frac{\pi q}{N_x N_t}}$.

The homogeneity of the action density is the aim of the construction \eqref{eq:insta_U2}, as it is a requirement for 2D configurations to be of locally minimal action.
Merely multiplying each link of \eqref{eq:insta_U1} with unity $\mathbf{1}$ fails to achieve this for arbitrary $q$.
It would amount to a configuration of minimal action for $q \in N_c\,\mathbb{Z}$, as the topological charge of this configuration has an additional factor of $N_c$ due to the determinant in \eqref{eq:q}.
The configuration thus obtained is included in \eqref{eq:insta_U2} by taking $|\vec{u}| = |\vec{v}| = 0$.
If one takes the $N_c$-th root of the $U(1)$-parts of the links, one does produce a configuration of topological charge $q$, but it is not of minimal action.
This is because the plaquette at $(x,t) = (N_x, N_t)$ differs from the rest by a $\mathbb{Z}_{N_c}$-factor, leading to an inhomogeneous action density.
For $N_c = 2$ our construction \eqref{eq:insta_U2} fixes this, but for $N_c\geq 3$ we have not succeeded yet in constructing instanton configurations for arbitrary $q\in\mathbb{Z}$.

\begin{figure}[!tb]
    \centering
    \subfigure{
        \includegraphics[width=0.5\linewidth]{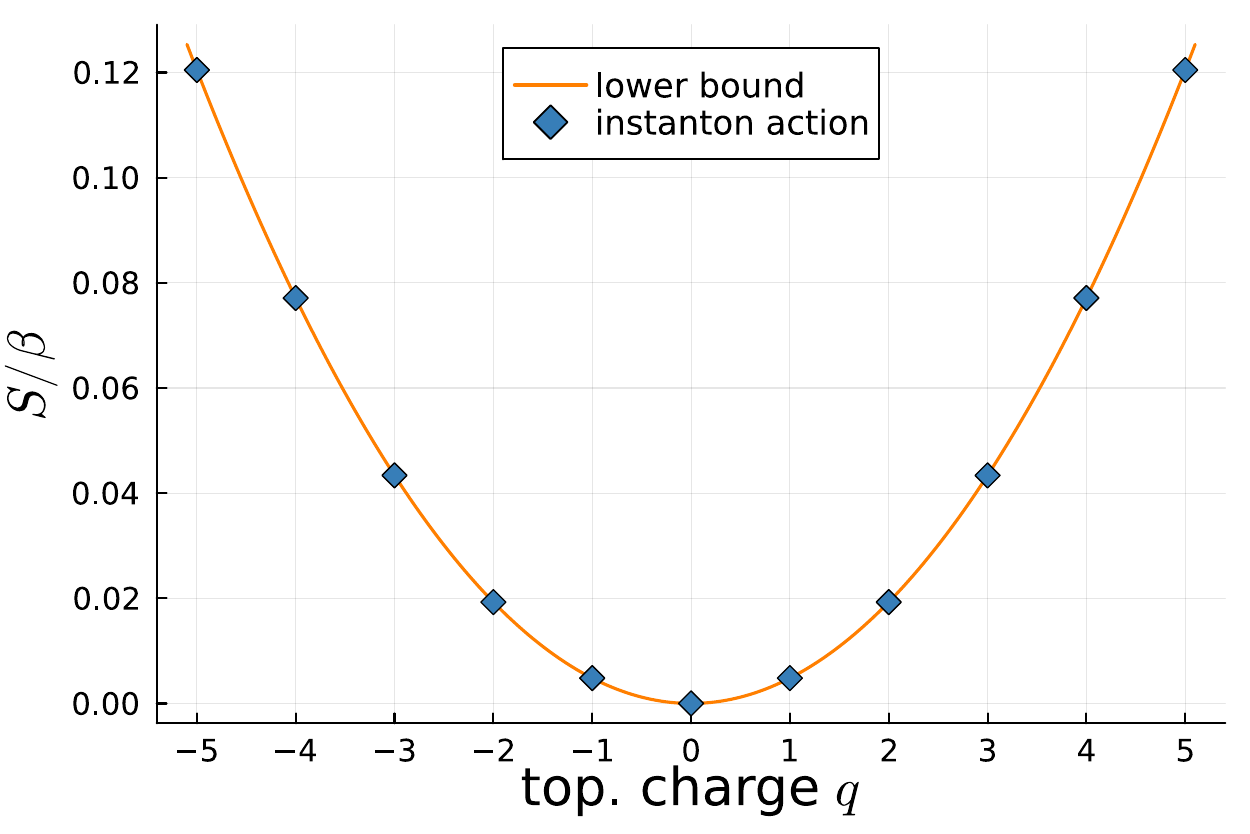}
    }%
    \subfigure{
        \includegraphics[width=0.5\linewidth]{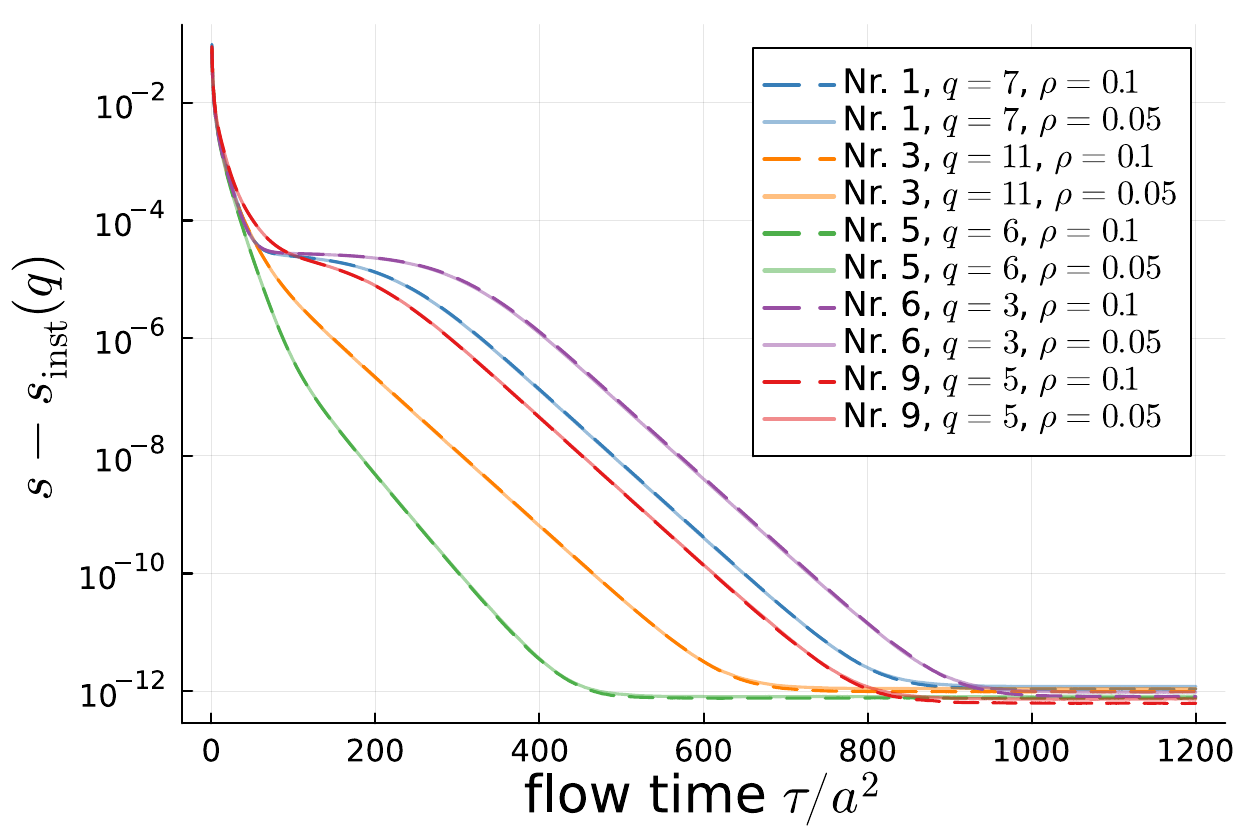}
    }%
    \caption{Left: Action of the $U(2)$ instanton configurations \eqref{eq:insta_U2} as well as the lower bound \eqref{eq:S}. Right: excess of the action density of five thermalized configurations ($\beta = 6.0$) over the bound \eqref{eq:S} plotted against the gradient flow time $\tau = \rho\cdot N_\text{stout}$ with two step sizes $\rho$. Square lattices of $L/a = 32$ are used in both cases. }
    \label{fig:bogo_smeared}
\end{figure}

It is interesting to test whether thermalized configurations would evolve towards such $q$-instanton configurations under gradient flow.
To that end consider the action of \eqref{eq:insta_U2}
\begin{equation}\label{eq:S}
    \frac{S^{U(2)}_\mathrm{inst}}{\beta} = N_x N_t \left(1 - \cos\left(\frac{\pi q}{N_x N_t} \right)\right),
\end{equation}
which is illustrated in Fig.\,\ref{fig:bogo_smeared} (left).
The right panel shows that the action of thermalized configurations with fixed $q$ converges to \eqref{eq:S} under gradient flow, which is in line with the fact that \eqref{eq:insta_U2} is of minimal action in a given topological sector.

For odd $q$ it can be shown that any two choices of $\vec{u}$ and $\vec{v}$ in \eqref{eq:insta_U2} are gauge equivalent. 
For even $q$ this is not the case, as $|\vec{u}|$ and $|\vec{v}|$ can be chosen freely, allowing for further transformations that do not change the action and cannot be represented by a gauge transformation.
Consider \eqref{eq:insta_U2} after transforming it into maximal tree gauge $U_t(x,t) = \mathbf{1} \,\forall t<N_t$ and $U_x(x,1) = \mathbf{1}\,\forall x<N_x$.
In this gauge, the last $t$-slice $\{U_t(x,N_t) \,\big|\, x = 1...N_x \}$ and the last $x$-slice $\{U_x(N_x,t) \,\big|\, t = 1...N_t \}$ can always be multiplied by a center element of the gauge group (here $U(1)$).
For \eqref{eq:insta_U2} with even $q$, however, a multiplication by $\exp(\mathrm{i}r\vec{u}\vec{\sigma})$ resp. $\exp(\mathrm{i}s\vec{v}\vec{\sigma})$ with suitable $r,s\in\mathbb{R}$ is also possible.

\section{Special topological configurations}

\begin{figure}[!tb]   
    \centering
    \includegraphics[width=0.9\linewidth]{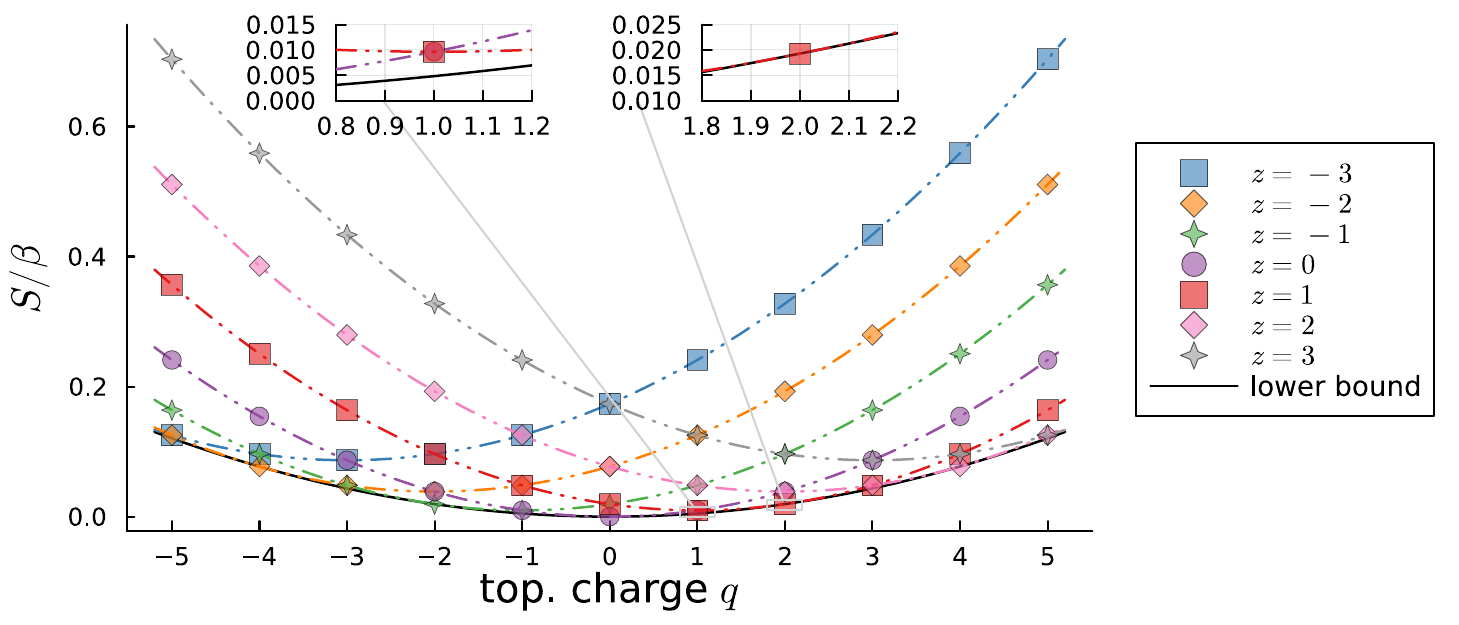}
    \caption{The action of \eqref{eq:local_min}, evaluated for various $(q,z)$-combinations ($N_c = 2$). The lower bound is Eq.\,\eqref{eq:S}. }
    \label{fig:local_inst}
\end{figure}

For 2D $U(N_c)$ theory one may derive further configurations of homogeneous action density
\begin{subequations}\label{eq:local_min}
    \begin{align}
        U_x(x,t) &= \mathrm{e}^{-\mathrm{i}t \frac{2 \pi q}{N_c} \frac{1}{N_x N_t}} \, 
            \exp\left( -\mathrm{i}t\,\frac{2\pi}{N_c} \,\frac{1}{N_x N_t} \big(z N_c - q\big) \, \mathrm{diag}(1,...,1,1-N_c) \right), \\
        U_t(x,t) &= \mathrm{e}^{\mathrm{i}x \frac{2 \pi q}{N_c} \frac{1}{N_x} \ \delta_{t,N_t}} \, 
            \exp\left( \mathrm{i}x\,\frac{2\pi}{N_c} \,\frac{1}{N_x} \big(z N_c - q\big) \, \delta_{t,N_t} \, \mathrm{diag}(1,...,1,1-N_c) \right),
    \end{align}
\end{subequations}
where $z\in\mathbb{Z}$, and we have plugged in the last generator $\lambda_{N_c^2-1}/2$ of the Lie algebra $\mathfrak{su}(N_c)$ with
\begin{equation}
    \lambda_{N_c^2-1} = \sqrt{\frac{2}{N_c^2-N_c}} \, \mathrm{diag}(1,...,1,1-N_c)\, .
\end{equation}
Hence, for a given charge $q$ there is an infinite tower of such configurations \eqref{eq:local_min}, and we find
\begin{subequations}
\begin{align}\label{eq:plaq}
    U_\Box(n)\Big|_{n = (N_x,N_t)} &= \mathrm{e}^{\mathrm{i}\frac{2\pi q}{N_c} \left(\frac{1}{N_x N_t} -1\right)}
        \exp\left(\mathrm{i}  \frac{2\pi}{N_c}  \left(\frac{1}{N_x N_t} -1 \right) \big(z N_c - q \big)  \mathrm{diag}(1,...,1,1-N_c) \right), \\
    U_\Box(n)\Big|_{n \neq (N_x,N_t)} &= \mathrm{e}^{\mathrm{i}\frac{2\pi q}{N_c} \frac{1}{N_x N_t}}
        \exp\left(\mathrm{i}  \frac{2\pi}{N_c}  \frac{1}{N_x N_t} \big(z N_c - q\big)  \mathrm{diag}(1,...,1,1-N_c) \right),\\
    \mathrm{Re\, Tr}\, U_\Box(n) &= (N_c-1)\cos\left(\frac{2\pi z}{N_x N_t} \right) + 
        \cos\left(\frac{2\pi}{N_x N_t} \Big(q-(N_c-1)z\Big)\right). \label{eq:ReTrP}
\end{align}
\end{subequations}
The untraced plaquette $U_\Box(n)$ depends on the site $n$, while $\mathrm{Re\, Tr}\, U_\Box(n)$ is independent of the location.
Hence, for any $N_c$ the action density of a ``special configuration'' \eqref{eq:local_min} depends only on $q$ and $z$. 
Some of them (for $N_c = 2$ and a $32^2$ lattice) are shown in Fig.\,\ref{fig:local_inst}.

For $N_c = 2$ the ``special configurations'' \eqref{eq:local_min} contain, as a subset, some of the instanton configurations discussed in Sec.\,\ref{sec:global}, but not all of them. 
For instance the choice $(q,z) = (2,1)$ in \eqref{eq:local_min} yields the same configuration as $q = 2, \vec{u}=\vec{v}=0$ in \eqref{eq:insta_U2}. Accordingly, this makes \eqref{eq:ReTrP} assume the value \, $2\cos\left(\frac{2\pi}{N_x N_t}\right)$, which matches \eqref{eq:S} in the case $q=2$. 
But for $q=1$ there is no choice of $z$ which would match the lower bound \eqref{eq:S}. 
This situation is illustrated in Fig. \ref{fig:local_inst}.
For arbitrary $N_c$ we recover the instanton configurations for $q\in N_c\mathbb{Z}$\, by dialing $z=q/N_c$, but the ``special configurations'' \eqref{eq:local_min} do not provide any help in constructing $q$-instanton configurations for $q\notin N_c\mathbb{Z}$.

\begin{figure}[!tb]
    \centering
    \includegraphics[width=0.6\linewidth]{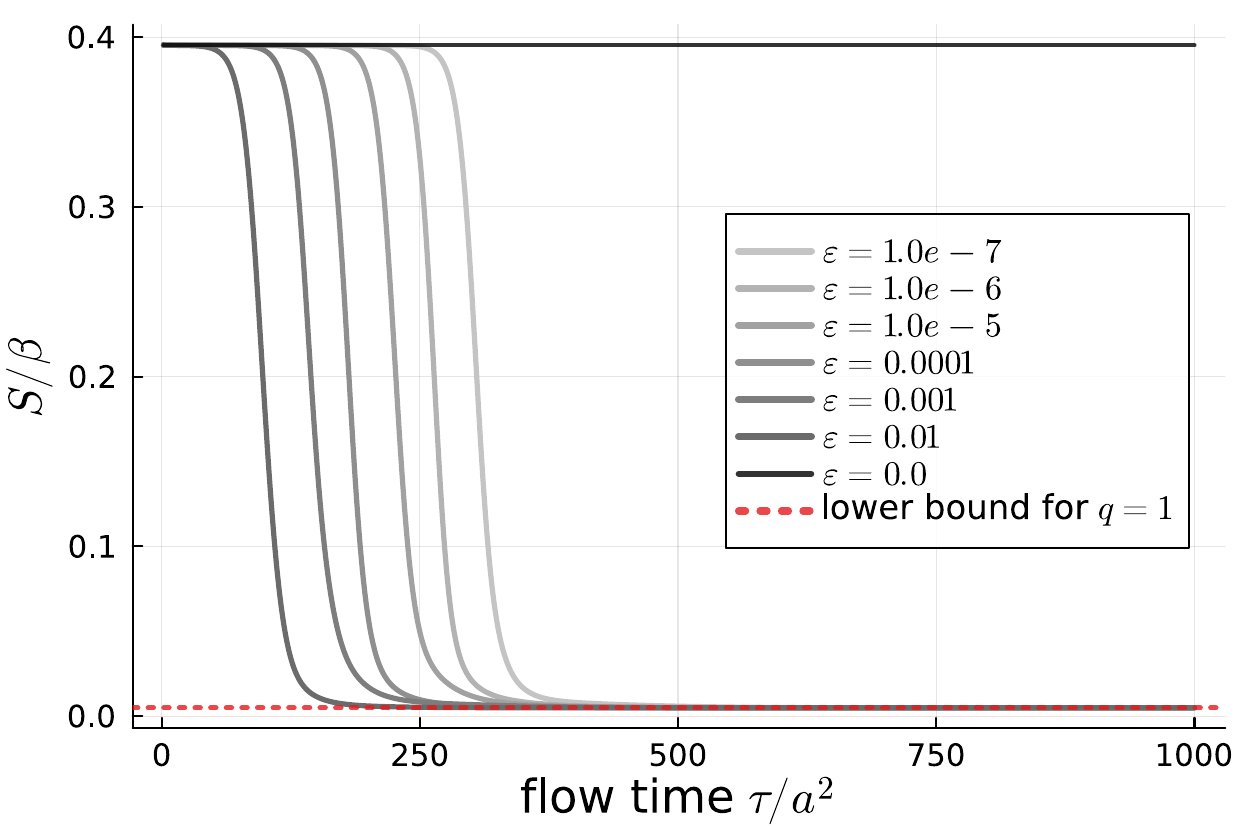}
    \caption{The action of the ``special configuration'' \eqref{eq:local_min} with $N_c = 2$ and $(q,z) = (1,5)$ under gradient flow, after each link has been multiplied with a random $U(2)$-element of step size $\varepsilon$, for $N_x = N_t = 32$. The lower bound is Eq.\,\eqref{eq:S}. Without perturbation the configuration \eqref{eq:local_min} seems to be stable under gradient flow. }
    \label{fig:disturbed_locals}
\end{figure}

When each link of a ``special configuration'' \eqref{eq:local_min} for $N_c = 2$ is multiplied by a random $U(2)$-element of step size $\varepsilon$ and the result subject to gradient flow, the action decreases until it reaches the instanton action \eqref{eq:S} of the corresponding $q$-sector.
For smaller $\varepsilon$ the decrease of the action occurs later in the course of the gradient flow, and for $\varepsilon = 0$ no decrease is observed at all, see Fig.\,\ref{fig:disturbed_locals}.

\begin{figure}[!tb]
    \centering
    \subfigure{
        \includegraphics[width=0.5\linewidth]{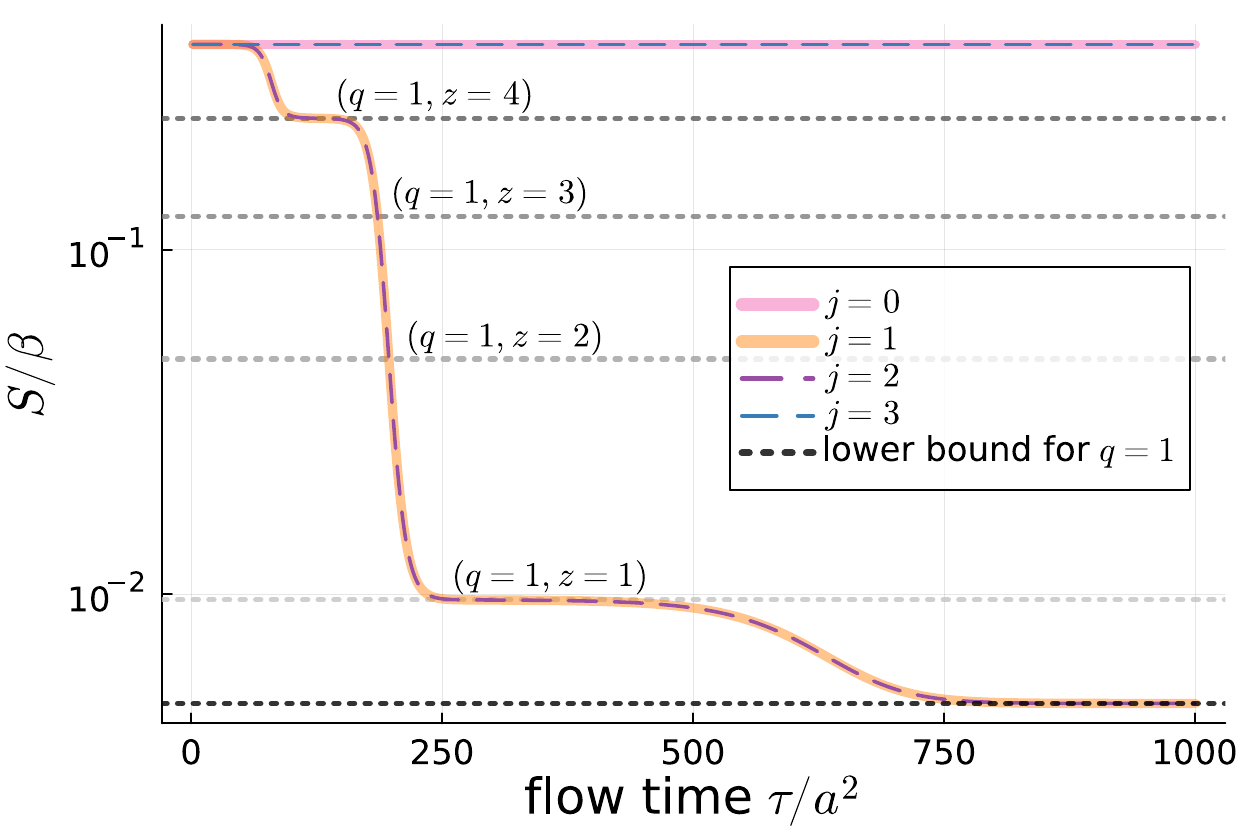}
        \label{fig:locally_dist_locals_a}
    }%
    \subfigure{
        \includegraphics[width=0.5\linewidth]{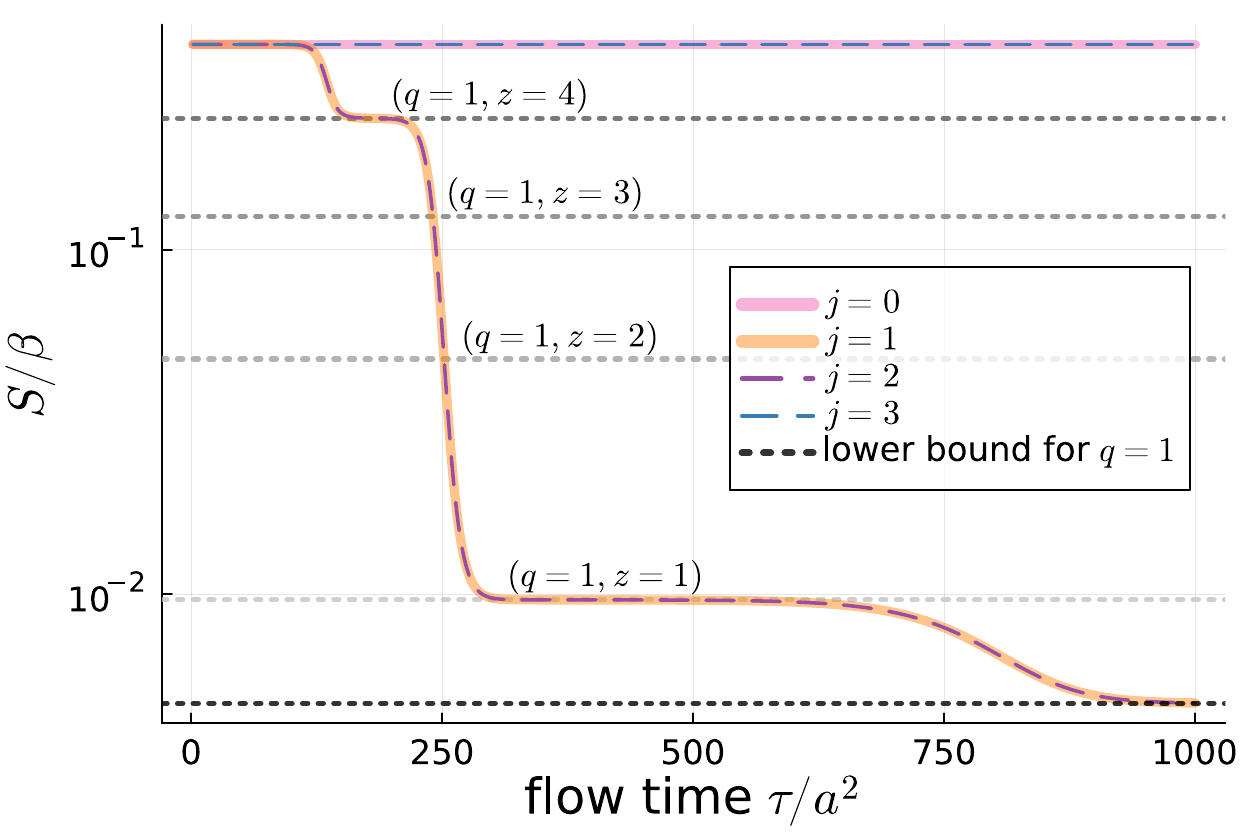}
        \label{fig:locally_dist_locals_b}
    }%
    \caption{The action of the ``special configuration'' \eqref{eq:local_min} with $N_x = N_t = 32$, $N_c = 2$ and $(q,z) = (1,5)$ under gradient flow, after an arbitrary link has been multiplied with a $U(2)$-element of the form $\exp(\mathrm{i}\varepsilon\sigma_j/2)$. The grey, dotted lines show the action of \eqref{eq:local_min} for $q=1$ and various values of $z$. Left: $\varepsilon = 10^{-1}$, right: $\varepsilon = 10^{-3}$. }
    \label{fig:locally_dist_locals}
\end{figure}

If only one link is perturbed by a small $U(2)$-element of the form $\exp(\mathrm{i}\varepsilon\sigma_j/2)$ the behavior depends on the Lie direction $j$.
For $j\in\{0,3\}$ (where $\sigma_0$ is the identity) the action remains unchanged under subsequent gradient flow, but for $j\in\{1,2\}$ the action decreases, see Fig.\,\ref{fig:locally_dist_locals}.
Contrary to the case of a global perturbation as in Fig.\,\ref{fig:disturbed_locals}, the action forms intermediate plateaus at the action levels of some ``special configurations'' whose actions lie in between that of the original configuration and the global minimum \eqref{eq:insta_U2} of the $q$-sector. 
This behavior is found for any position of the perturbed link and has been checked for $(q,z)$ combinations with $q = 0,...,5$ and $z = 0,...,3$.
The one feature which we find repeated is that a smaller perturbation $\varepsilon$ leads to a longer relaxation time $\tau/a^2$, but the final configuration is always the global minimum \eqref{eq:insta_U2} of the respective $q$-sector, see Fig.\,\ref{fig:locally_dist_locals}.

\section{Open questions}

\begin{figure}[!b]
    \centering
    \subfigure{
        \includegraphics[width=0.5\linewidth]{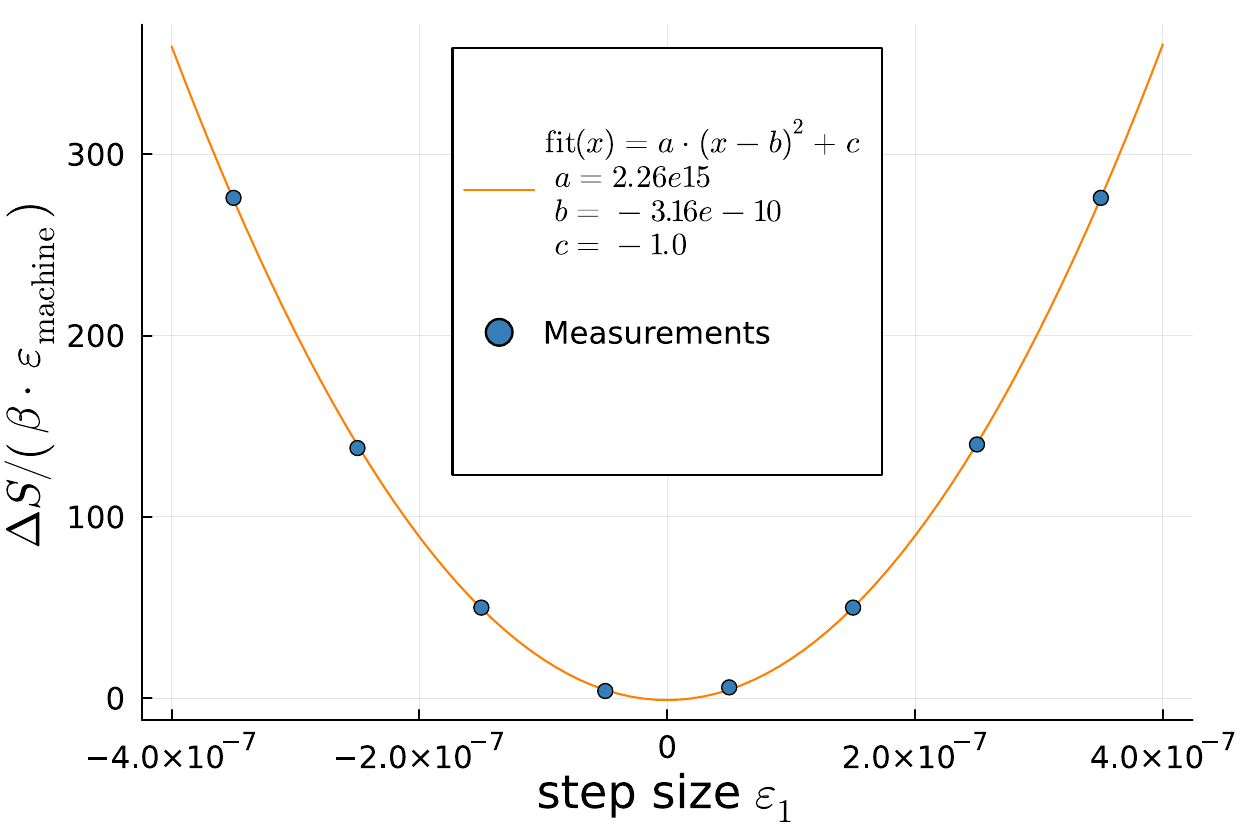}
    }%
    \subfigure{
        \includegraphics[width=0.5\linewidth]{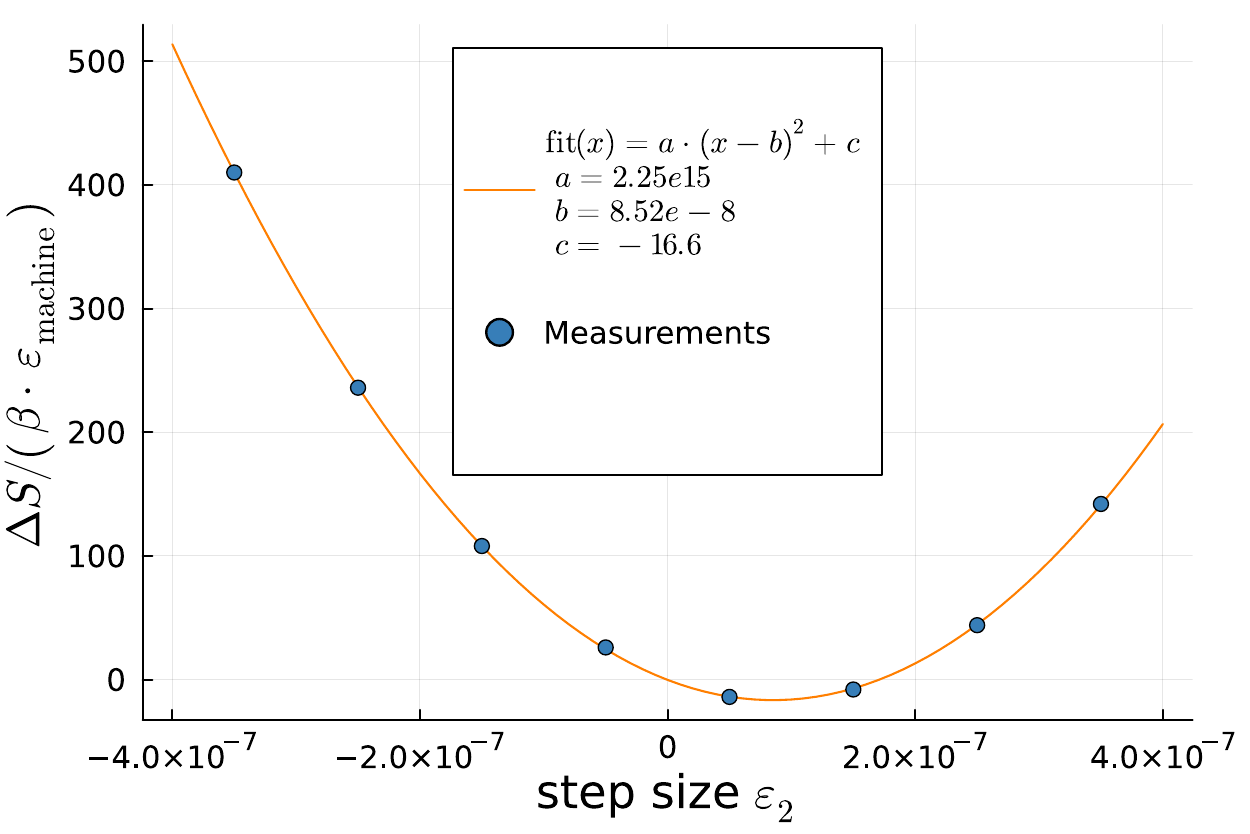}
    }%
    \caption{Left: a link of a ``special configuration'' \eqref{eq:local_min} with $N_x = N_t = 32$, $N_c = 2$ and $(q,z) = (1,5)$ is perturbed by a $U(2)$-element of step size $\varepsilon_1$ and $\Delta S$ is measured in units of machine precision $\varepsilon_\mathrm{machine}$. Right: after a perturbation $\varepsilon_1 = 10^{-3}$ along an algebra direction $j_1$, a perturbation $\varepsilon_2$ along $j_2\perp j_1$ is applied. }
    \label{fig:sdiff}
\end{figure}

In the previous section we found that perturbing an arbitrary link of a ``special configuration'' \eqref{eq:local_min} with a given $(q,z)$ combination in the $j=1$ or $j=2$ direction in color space suffices to make the result flow into the global minimum of the respective charge $q$ sector. 
But after a perturbation in the $j=0$ or $j=3$ direction it is stable under gradient flow (as is the unaltered configuration).

Given this dichotomy, the question arises whether a member of the family \eqref{eq:local_min} of uniform-action configurations (dubbed ``special configurations'' above) is a local minimum of the action in the space of gauge configurations or instead a saddle point. 
Since, as of this writing, we do not know the answer, we briefly mention a few steps taken so far to clarify the issue.

In the first place one may perturb an arbitrary link $U_\mu(n)$ of \eqref{eq:local_min} with a $U(N_c)$ element $\exp(\mathrm{i}\varepsilon\lambda_j/2)$ and measure the action difference $\Delta S$ for a few step sizes $\varepsilon$. 
One expects to find a parabola with a minimum at $\varepsilon=0$. 
Doing this for $N_c=2$ we find this expectation confirmed, see Fig.\,\ref{fig:sdiff} (left). 
But we were surprised to see that the parabola has universal features -- it's always the same parabola, regardless of $(q,z)$, the position $n$, direction $\mu$ and the Lie direction $j$ (also replacing $\sigma_j$ by $r_0\sigma_0+\vec{r}\vec{\sigma}$ with $||r||=\varepsilon$ leaves it unchanged).

Still, this observation does not explain why a small perturbation of a single link of a ``special configuration'' \eqref{eq:local_min} may suffice to make the result evolve, under gradient flow, towards the sector minimum \eqref{eq:insta_U2} of that $q$ value. 
Therefore we decided to perturb an arbitrary link in Lie direction $j_1$ with a step size $\varepsilon_1=0.001$, followed by a second perturbation $\varepsilon_2$ of the same link in an orthogonal direction $j_2 \neq j_1$. 
This second $\Delta S$ curve is no longer universal; for some $(j_1,j_2)$ combinations its minimum is at $\varepsilon_2 \neq 0$, see Fig.\,\ref{fig:sdiff} (right). 
But it is worth mentioning that the decrease in the second step is orders of magnitude smaller than the increase in the first step.

Altogether these tests nurture the belief that any ``special configuration'' \eqref{eq:local_min} is a local minimum of the action in the $U(N_c)$ theory, but more research is needed to establish this firmly.

\section{Conclusion}

We have explored a few topics in $U(N_c)$ gauge theories in 2D with periodic boundary conditions. 
These theories are interesting, since they have topological features akin to $SU(N_c)$ theories in 4D.

Analytical expressions for the average plaquette and the topological susceptibility agree with our Monte Carlo (MC) data, but they suggest that it takes rather high $\beta$-values to enter the Symanzik scaling regime for these quantities.

We were able to write down exact $q$-instanton configurations (7) in the case of $N_c=2$, that is gauge fields $U_\mu(n)$ with topological charge $q$ and minimal Wilson action $S$ in that sector. Unlike instantons in 4D, they have uniform action density.

Given a configuration $U_\mu(n)$ in a thermalized MC stream, the gradient flow lets it evolve towards a gauge transformed version of the minimum solution (7) in that $q$-sector. 
In 2D the gradient flow is thus ``trivializing'' within a given topological sector $q$, but not globally.

Finally, we constructed a class of ``special configurations'' (9) of homogeneous action density for arbitrary $N_c$, labeled by a pair $(q,z)\in\mathbb{Z}^2$. 
They seem to play a special role in 2D $U(N_c)$ theories, but more work is needed to elucidate their role.

\section*{Acknowledgements}

\noindent We would like to thank Christian Hölbling and Lukas Varnhorst for helpful discussions.

\end{document}